\documentclass[preprint]{aastex}

\def\<#1>{\langle#1\rangle}
\def\pcite#1{[ref]}

\def\Msun{M$_\odot$}
\def\ds{\displaystyle}
 
\slugcomment{Revised: \today}
 
\shortauthors{Dalia Chakrabarty \& Simon Portegies Zwart}
\shorttitle{Studying Clusters Inversely}
 
\begin{document}    

\title{An Inverse Problem Approach to Cluster Dynamics}

\author{Dalia Chakrabarty\altaffilmark{1}}
\affil{Department of Physics \& Astronomy\\ 
Rutgers University\\
136 Frelinghuysen Road\\
Piscataway, NJ 08854-8019, USA}

\author{Simon Portegies Zwart}
\affil{Astronomical Institute 'Anton Pannekoek' and 
Dept. of Computer Science\\
University of Amsterdam\\
Kruislaan 403 1098SJ \\
Amsterdam, the Netherlands}

\begin{abstract}
We propose a new non-parametric algorithm that can be implemented to
study and characterize stellar clusters. The scheme attempts to
simultaneously recover the stellar distribution function and the
cluster potential by using projected radii and velocity information
about the cluster members. The pair of these functions that is most
consistent with the input data is detected by the Metropolis
algorithm. In this work, the cluster characteristics recovered by
CHASSIS are calibrated against the N-body realizations of two
clusters, namely Hyades and Arches. The cluster mass and line-of-sight
projected velocity dispersion profiles are correctly reproduced by the
algorithm when the cluster obeys the assumption used in the code,
namely isotropy in phase space.  The results recovered by the code are
shown to be insensitive to the choice of the initial parameters. The
results are also not influenced by increasing the number of input data
points as long as this number exceeds a minimum value which is
moderately low for an input data set that obeys the assumptions of
isotropy and sphericity.
\end{abstract}


\section{Introduction}

\noindent
The characterization of stellar clusters is more often than not, an
exercise in model building. One usually begins with an assumption (of
sphericity) for the cluster geometry. Quite often, luminosity
measurements are used in conjunction with fitting algorithms to
provide the surface brightness profile of the cluster
\citep{larsen01}. However, a major shortcoming of such a procedure is
that such fits are often unstable over the whole radial range
considered. Moreover, the robustness of these fitting techniques is
always challenged by details in the observed data.  It is possible in
principle to obtain smooth approximations to the surface density
profile and projected velocity dispersion profiles from
observations. These projected quantities could be inverted to yield
their total or three dimensional counterparts via the Abel integral
equation; such a deprojection is unique in the spherical case.
Unfortunately an inversion of this nature is highly sensitive to shot
noise in the data bins. Genzel et al. (1996) acknowledge this
problem and adopt the more secure alternative of starting with
parametrized forms of the density and radial velocity dispersion
profiles, which are projected back to observational space, while
maintaining the search for those projected profiles that fit the
observed result best. Such an inverse approach, whenever possible, is
almost always the better option since it involves less
errors. Frequently, a cluster is considered to be in virial
equilibrium; velocity dispersion estimates or enclosed mass
information are then extracted from the virial equation
\citep{bosch01, ghez}. A method that can substitute this assumption of
a virialized cluster is certainly more satisfactory since observations
alone cannot be sufficient to establish the validity of this
conjecture.

This state of affairs can be improved upon by considering the inverse
problem approach. It is indeed true that such a formalism demands a
moderately large data set. Kinematic information about the members of
a number of observed stellar clusters suffices to fulfill this
requirement. In this paper, we propose an non-parametric technique
which can be used for the characterization of stellar systems. The
algorithm presented here has been used before \citep{dalpras} for
the estimation of the mass enclosed in the inner few parsecs of the
Milky Way (hereafter Paper~I). We refer to our scheme as CHASSIS, i.e.
CHAracterization of Stellar Systems using a new Inverse Scheme.

This paper aims to advance the case of using CHASSIS in investigations
of stellar clusters. In order to establish the resourcefulness of this
code, we test our inverse algorithm on two simulated star clusters,
which were placed at known distances and projected on a selected part
of phase space. We present the results of analyzing these simulated
star clusters with CHASSIS and then compare the results with the
cluster simulations. This comparative exercise reflects the degree of
robustness of the algorithm.

\section{Initial conditions and calculations}
\label{sec:N-body}

\noindent
As input for the simulations we adopt two clusters which we use as
templates for the initial conditions. We call these models A
(mimicking the Arches star cluster) and H (mimicking the Hyades star
cluster, see Table~\ref{tab:N-body}).

The cluster simulations were performed in three steps:
1) initialization of the stellar system, 2) calculation of the
evolution of the cluster to a certain time and 3) preparation of the
simulated star cluster to use as input for the inverse algorithm.

Details about the Hyades (model H) and the global setup of the
calculations of the Arches model (model A) are published
\citep{simon01, simon02}.  One ingredient of the Arches model that
makes it different compared to previously performed N-body
calculations is the inclusion of the external tidal field of the
Galaxy and the frictional drag force on the bound cluster. As a
result, the cluster spirals in from an initial distance of 6pc from
the Galactic center to less than 2pc, in about a million years. The
strong tidal field in model A causes this cluster to become strongly
aspherical and elongated perpendicular to the line connecting the
cluster with the Galactic center. However, CHASSIS works on the
assumption of sphericity of the cluster, as is explained below. In the
light of this obvious contradiction, the results obtained from the
inverse algorithm for the Arches cluster were expected to be
interesting. Model A was run on a GRAPE-6 and took about 188 CPU
hours.  Model H took about 55 hours on a single board GRAPE-4.

\begin{table*}
\caption{Initial and final conditions for the N-body models of the Arches
 and Hyades clusters. Columns give the initial density profile (King
1966 model parameter), the number of stars, the total cluster mass
(in\, \Msun), the virial radius (in parsec) and the tidal radius of
the cluster and its distance to the Galactic center. The subsequent
columns give the parameters for these models at the moment the data
was analyzed by the inverse algorithm; the time (in Myr), the number
of remaining stars bound to the cluster, the total mass of these stars
(in \Msun), the half mass radius and the distance to the Galactic
center.}
\vspace{1cm}
\centerline
{
\begin{tabular}{cc|cccccccccc} \tableline
Name& Wo& N& M& Rhm& Rt& Rgc& T& N& M& Rhm& RGC \\
{}&{}&{}&(\Msun)&(pc)&(pc)&(pc)&(Myr)&{}&(\Msun)&(pc)&(pc) \\ \hline
A   & 3& 64k& 62600& 0.17& 0.52&6.0& 0.97& 55224& 54200&  0.093& 2.8 \\
H   & 6&  3k&  1606& 2.50& 13.7&13k&  400&  2628&  1192&   4.66& 13k \\
\tableline
\end{tabular} 
}
\label{tab:N-body} 
\end{table*}

For the selected models, positions and velocities of all stars are
calculated self consistently together with the evolution of the stars
and binaries in the clusters. Calculations are performed using the
starlab software environment (see {\tt http://manybody.org}) and
utilize the GRAPE-4 and GRAPE-6 special purpose computers
\citep{makino1997}. The simulations were stopped and ported to the
inverse algorithm when the age of the cluster is about 1 million years
for model A and about 400 million years for model H. For model H all
the stars remaining in the cluster after 400 million years were ported
to the inverse algorithm while for model A the stars remaining after 1
million years were used. From these sets various smaller data sets
were selected for direct input into CHASSIS (see below). 

\subsection{Preparation of the clusters for the inverse algorithm}

\noindent
The stellar samples used as inputs in our algorithm are selected from
the N-body data sets that describe the two clusters. We use kinematic
data of a maximum of about 2600 stars belonging to Hyades in our
work. The number of stars in Arches from which input data is chosen is
much higher (about 55,000). The input files for the algorithm are
prepared by choosing stars either at a random or by picking stars
according to their luminosity. The kinematic information provided in
the N-body data sets about each star is made to undergo appropriate
transformations in order to calculate the corresponding apparent
position, radial velocity and transverse velocity. These quantities
are the inputs for our code CHASSIS. Diagrammatic representations of
the kinematics of the two clusters are in Figure~\ref{fig:data200} and
Figure~\ref{fig:data64}. These are plots of radial and transverse
velocities against radial positions of the stars on the plane of the
sky, i.e. the apparent position of the stars.

The N-body simulations indicated that the clusters are neither
isotropic, nor are they perfectly spherical. The extent of the
presence of anisotropy and asphericity in the A and H models is
brought out in Figure~\ref{fig:anisotropy} which displays the radial
dependence of the ratios of the $x$ and $y$ components of the velocity
dispersion vector ($\sigma_{vx}$ and $\sigma_{vy}$) to the $z$
component of the same ($\sigma_{vz}$) and the $x$ and $y$ component of
the spatial dispersion vector ($\sigma_x$ and $\sigma_y$) to the $z$
component of the same ($\sigma_z$). Such ratios are only crude
estimates of the deviation from sphericity and isotropy, but they
serve our purpose of estimating the degree to which we can expect the
input data to be incongruous with the assumptions of the code.

As can be seen from this figure, in the H-model, the ratios of the
spatial dispersions is about 1 to a radius about 5pc, after which both
$\sigma_{x}/\sigma_{z}$ and $\sigma_{y}/\sigma_{z}$ rise with radius
slowly though a marked discontinuity in the run of the spatial
dispersions shows up just inside a radius of about 10pc. A big jump
occurs in the ratios $\sigma_{vx}/\sigma_{vz}$ and
$\sigma_{vy}/\sigma_{vz}$ between about 4pc and 7pc. This picture
prompts us to choose to work within about 10pc in the Hyades cluster,
knowing that the kinematic distribution of this input data is not very
well consistent with isotropy in this radial range. 

The equivalent radial cutoff for the A-model appears to be about
0.2pc, though further on, the cluster gets highly aspherical and
anisotropic. So we choose to select kinematic information of the
A-model from radii within 0.2pc.; however when viewed at a finer
resolution, even within this radius, the run of the dispersion ratios
is quite jerky, indicating that the data from the Arches cluster does
not abide by the assumptions of isotropy in the code.

The chosen radial ranges are also inclusive of the half-mass radii of
the two clusters, about 4.66pc for Hyades and 0.093pc for Arches. From
the H-model and the A-model, the mass enclosed within 4.6pc and 0.093pc are
approximately 584M$_{\odot}$ and 24522M$_{\odot}$, respectively. These
mass values are very close to half the mass of the cluster (see
Table~\ref{tab:N-body}).

\section{Algorithm used}
\label{sec:code}
\noindent
The code used in our work was introduced and discussed in details in
Paper~I. Observed kinematic data (apparent positions of stars on the
plane of the sky and their velocities) is used as input in this
algorithm. CHASSIS is a non-parametric code which attempts to identify
both the equilibrium stellar distribution function (DF) from which the
observed kinematical data is drawn as well as the potential of the
observed stellar system. The code assumes sphericity of this stellar
system. In its present form the scheme also assumes an isotropic
DF. Paper~I reports a test that was developed to check the consistency of
the used data with the assumption of isotropy. The basic design of
our algorithm is described briefly below.

As an initial guess, the code requires a set of functions in the form
of an arbitrarily chosen trial potential ($\Phi$) (rather the
corresponding density profile $\rho$) and distribution function $f$,
which is a function of energy $E$ only, via the assumption of
isotropy. The final results have no bearing on the initial
configurations. In order to obtain a potential from a chosen density,
a discretized density profile is used, i.e. $\rho(r)$ is held a
constant over the radial bin defined around radius $r$. The
calculation of the potential of a series of spherical mass shells is
simple and is given in Paper~I. The only constraints placed on these
trial functions is that the density profile is monotonically
decreasing in radius and the DF is monotonically increasing in
effective energy. Adherence to these conditions is demanded every time
the profiles are altered to generate new profiles. Such tweaking of
the functions takes place at the end of each iterative step.

At every iterative step, the current DF is projected into observable
space, using the observed kinematical data. Thus, if we have
information on the line-of-sight (LOS) velocity $v_p$ and the apparent
position $r_p$, then:
\begin{equation}
\nu_p(r_p,v_p)=\ds{
		   \int{dz
	                \int\int{dv_xdv_yf[v_r^2+v_t^2+2{\Phi}(r)]
                             }
                       }}.
\label{eqn:proj}
\end{equation}
Here $r$ is the spherical radius and $r_p$ the cylindrical radius
on the plane of the sky, so that $r_p^2=x^2+y^2=r^2-z^2$; $v_r$ and
$v_t$ are components of velocity parallel and tangential to the radius
${\bf{r}}$. We choose the $y$-axis to lie completely in the
plane containing the radius vector ${\bf{r}}$ and the LOS. This implies
that $v_r^2=(v_y\sin\theta)^2+(v_z\cos\theta)^2$, $\theta$ being the
polar angle, i.e., $\cos\theta=z/r$.

This projection of $f(E)$ to $\nu_p(r_p,v_p)$ could be achieved if a
smooth approximation to $f$ is considered. Another even simpler way
(introduced in \citep{david}) used in our work is to hold $f$ a
constant over any integral cell defined around a given value of $E$. Then
corresponding to this energy bin, the projected distribution function
is
\begin{equation}
\nu_p^{\rm cell}(r_p,v_p)=\int{dz\, A(r,r_p,v_p)},
\end{equation}
where $A(r,r_p,v_p)=\int\int{dv_xdv_y}$ is the area that the energy
bin occupies in phase space. Estimation of this area requires
knowledge of the bounds on individual phase space coordinates that
correspond to every energy bin. Identification of these bounds is
discussed in Paper~I. The total projected distribution function is a
simple sum over all the energy bins, i.e.,
\begin{equation}
\nu_p(r_p,v_p)=\sum_{i}f_i\nu_{pi}^{\rm cell}(r_p,v_p).
\end{equation}
The product of all the projected distribution functions,
each obtained for a pair of $r_p$ and $v_p$ in the observed data-set,
gives a likelihood function $L$.
\begin{equation}
\log(L) = \sum_{\rm{i=1}}^{\rm{N}}\nu_{pi},
\end{equation}
where $N$ is the total number of pairs of data points in the set and
$\nu_{pi}$ corresponds to the $i^{\rm th}$ data point. 

When the accessible velocity data includes proper motion $v_{\mu}$,
(where $v_{\mu}=\sqrt{v_x^2+v_y^2}$) rather than LOS-velocity data,
the projection equation analogous to Eqn.~\ref{eqn:proj} is:
\begin{equation}
\nu_{\perp}(r_p, v_\mu) = \int{dz}\int{f(E)dv_z},
\end{equation}
while if the data includes complete velocity information, the projection
is simplest and is given by:
\begin{equation}
\nu_{\rm 3D}(r_p, v_\mu, v_p) = \int{dz\:f(E)}.
\label{eqn:3Dproj}
\end{equation}

The likelihood function is therefore a function in $n$-dimensions
where $n\:=\:n_r\times{n_E}$. Here $n_r$ is the number of radial bins
corresponding to the chosen density profile and $n_E$ the number of
energy bins characterizing the DF. The global maxima of the highly
non-linear function $L$ is sought in this $n$-dimensional space by the
Metropolis algorithm. A striking feature of using this method of
optimization is that we are ensured a whole array or ensemble of
models distributed according to the likelihood, rather than the single
model associated with the maximum likelihood. The availability of the
ensemble allows us to estimate uncertainties simply by measuring the
spread of values across the ensemble. This is a big advantage of using the
Metropolis algorithm. The maximization scheme is elaborated upon in Paper~I.

Thus in CHASSIS, at the $i^{\rm th}$ iterative step, the DF and
density profile are slightly tweaked from their forms used during the
$i-1^{\rm th}$ step. The choice of these functions for the $i^{\rm
th}$ step is referred to as ${\bf x_i}$. The current $f$ is projected
into observable space, at the currently chosen potential for each data
point, and the likelihood for this step $L({\bf x}_i)$ is
calculated. At the beginning Metropolis is expected to wander around
in the multi-dimensional domain of the likelihood function and $L$
tends to increase during this period. Soon, Metropolis settles down to
an ``equilibrium'' phase during which the likelihood is likely to be
close to the optimal value. Convergence is characterized in our code
by the step $k$ at which $L({\bf x}_k) <L({\bf x}_{k-N_{\rm
iter}})$. Here $N_{\rm iter} +1$ is the total number of iteration
steps that we use in any run. We recorded ${{\bf x}_{k-N_{\rm
iter}},\ldots,{\bf x}_k}$ as our likelihood-weighted sample of $\rho$
and $f$.

From the recorded DF and density profiles, mass within a pre-fixed
radius r (M(r)) is calculated as follows:
\begin{equation}
M(r) = \ds{\int_0^r{\rho(r)4\pi{r^2}dr}}
\end{equation}
In the discretized form, this equation is represented as:
\begin{equation}
M(r) = \displaystyle{
                     \frac{4\pi}{3}\left[
                     \delta^3\sum_0^j{\rho_i[3i^2 -3i +1]} +
                     \rho_j(r^3-j^3)\right]
		    }
\label{eqn:encmass}
\end{equation}
where $r$ lies in the range [$j\delta$, $(j+1)\delta$], i.e. $j =
Integer(r/\delta)$. $\delta$ is the width of a radial bin and
$j\in{N}$ such that $j\le{Integer}(r_{max}/\delta$). \\
The velocity dispersion profile $\sigma{(r)}$ is calculated from the
moments of $f(E)$.
\begin{equation}
\sigma{(r)}^2 = \ds{\frac{\int{f(E)\:v^2\:d^{3}{\bf v}}}
		         {\int{f(E)\:d^{3}{\bf v}}}} -
		\ds{\left(
			 \frac{\int{f(E)\:v\:d^{3}{\bf v}}}
			      {\int{f(E)\:d^{3}{\bf v}}}
		   \right)^2}
\label{eqn:DF_disp}
\end{equation}
where $v^2=v_p^2+v_\mu^2$ and $E=v^2/2 \:+\:\Phi(r)$. Thus, we obtain
a value of $\sigma(r_p)$ for every $r_p$ in the input data-set. 
This equation can be molded into the following difference equation:
\begin{eqnarray}
\sigma{(r)}^2 &=& \ds{2\left[\frac{3\sum_0^j{f_i\gamma_i^{5/2}(5i^4 -10i^3 +10i^2
-5i +1)}-3f_j(\Phi(r)^{5/2} - \gamma_j^{5/2})}
{5\sum_0^j{f_i\gamma_i^{3/2}(3i^2 -3i +1)}-3f_j(\Phi(r)^{3/2} - \gamma_j^{3/2})}\right]}- \nonumber\\ 
&& \ds{2\left[\frac
{3\sum_0^j{f_i\gamma_i^{2}(4i^3 -6i^2 +4i^2-1)}
-3f_j(\Phi(r)^{2} - \gamma_j^{2})}
{4\sum_0^j{f_i\gamma_i^{3/2}(3i^2 -3i +1)}
-4f_j(\Phi(r)^{3/2} - \gamma_j^{3/2})}
\right]^2}
\end{eqnarray}
where
\begin{equation}
\gamma_i = \ds{\frac{\delta}{i} - \frac{\Phi}{i^2}}
\end{equation}
Here, the current energy is $E_j$ that lies in the energy cell
$[j\delta, \:(j+1)\delta]$ i.e. $j=Integer(E/\delta)$. $\delta$ is the
width of an energy cell. Also, $j\in{N}$ such that $j
\le{E_{max}}/{\delta}$, where the maximum value of energy used in the
binning is $E_{max}$.

The error bars on the DF (and the density) are calculated at each
energy-bin (or radial bin) as the difference between the DF (or
$\rho$) at the $84\%$ level and that at the $16\%$ level. However, the
error bar on the value of velocity dispersion at any $r_p$ is the
difference between the value of the dispersion corresponding to the
density at the $84\%$ level and the value of $\sigma{(r_p)}$ estimated
with $\rho{(r_p)}$ at the $16\%$ level. This needs to be kept in mind
when the dispersion results are examined.

The dispersion profile obtained in this manner is subsequently
projected along the LOS and convolved with the density
distribution to yield the LOS projected velocity dispersion
profile which is a more realistic quantity from the point of view of
observations. This projected velocity dispersion is then compared to
the N-body profile.

\section{Implementation Details}
\noindent
The N-body realizations of the two clusters (Hyades and Arches) are
used as inputs in CHASSIS to obtain the cluster density and DF of the
constituent stars. The discretized density distribution is used to
derive the cluster mass profile (Equation~\ref{eqn:encmass}) and the
DF recovered by the code is transformed to give us the velocity
dispersion profile (see Equation~\ref{eqn:DF_disp}). In the following
sections, the enclosed mass and dispersion profiles of the N-body
clusters are compared to those estimated by the inverse algorithm.

Testing of the code is reported in Paper~I; the density and DF of a
set of stars described by the Plummer model were calculated by the
code. These quantities were found to match the Plummer density profile
and DF quite well, within error bars.

Another important facet of any inverse algorithm is to confirm the
independence of the results from the choice of the initial guess
(which comprises of an arbitrarily chosen pair of DF and density
profiles). We used very simple forms of density and distribution
function as the initial guess in the algorithm. The initial density
function is given by:
\begin{equation}
\rho(r) = \rho_0{\left(1 + \frac{r^2}{3r_c^2}\right)}^{\alpha},
\end{equation}
where $\rho_0$ is maintained a constant at 5000 and different choices
of $r_c$ and $\alpha$ determine different initial density
profiles. Changing $\rho_0$ was found to have no bearing on the
results while the results appeared most sensitive to the choice of
$\alpha$. The effect of changing the initial density
profiles was found to diminish with an increased number of iterative
steps $N_{\rm iter}$. At $N_{\rm iter}$=4000, density profiles defined
by $\alpha$ lying in the range [-0.3,-3.5] and $r_c\in{[1,7.5]}$, were
found to imply similar results. 

The initial DF has a normalization of unity and is chosen to have a
simple power-law form.
\begin{equation}
f(\Large{\varepsilon}) = (C + \Large{\varepsilon})^{\beta},
\end{equation}
where $\Large{\varepsilon}$ is the effective energy. A number of trial
runs with different values of the constant $C$ showed that the results
are independent of it even with a relatively low $N_{\rm iter}$. Thus
we used $C=0$ in all our main runs. $\beta$ on the other hand was a
parameter that had greater effect on the results. We used initial DFs
characterized by values of $\beta\in$ [1, 10.5]. Any effect of the
steepness of the initial choice of the DF on the results could be
negated by using a large enough $N_{\rm iter}$ (4000 was found to
suffice in the sense that the results were found to be consistent
within the error bars).

\section{Results from model H}
\noindent
The cluster properties that CHASSIS directly provides are the density
and the equilibrium DF of the stellar sample selected from the N-body
cluster. These cluster parameters are shown for the Hyades cluster in
Figure~\ref{fig:dfden_H}. Since this data includes information on all
three components of the velocity of each star, we use both LOS
velocity as well as proper motion in our algorithm. Thus, the
appropriate projection equation is given in Eqn.~\ref{eqn:3Dproj}. The
results obtained for this data-set are discussed below.

\subsection{Independence from Initial Model}
\label{sec:independence}
\noindent
In this section we present evidence to establish the insensitivity of
the results to the initial guess used in the algorithm. This is
displayed via the concurrence of the enclosed mass and velocity
dispersion profiles of the Hyades cluster, recovered by runs done with
different initial guesses. For all these runs the input to the
code constitutes of 37 stars, selected by their luminosity, from model H.

From Figure~\ref{fig:alfexp_massvel}, we can see that changing the
initial configuration results in profiles which agree with each other
within error bars. This offers confidence about the robustness of the
code insofar as changing the initial guess is concerned.
\subsection{Number of data points}
\label{sec:number}
\noindent
We conducted a series of numerical experiments to study the effect
that the number of data points, used as inputs to the algorithm, has
on the results suggested by the code. Samples of different sizes were
selected by their luminosity, from the data-set corresponding to the
N-body calculations that describe the Hyades cluster. The mass and
velocity dispersion profiles of the cluster were calculated for these
different input files and were compared to each other as well as with
their N-body counterparts. One quantity that is of interest is the
mass within the half-mass radius of the cluster. This is of course
derived from the enclosed mass profile but being a simple scalar
number, it can aid in gauging the extent to which the size of the
input file is effective in determining results.  The effect is shown
in Figure~\ref{fig:mass_hyades}. In addition to the value of the
enclosed mass from different runs, the figure also depicts how well
the predicted mass estimates tally with the N-body value.

The cluster profiles are shown in Figure~\ref{fig:ranL_hyades}.

The Figures~\ref{fig:mass_hyades} and ~\ref{fig:ranL_hyades} show that
the number of stars for which kinematic information is fed into the
algorithm, is not particularly influential in determining the mass
profile accurately. It is however true in general that the errors in
the calculation of the cluster parameters that are directly recovered
by the code reduce with an increase in the input file size. This is
discussed in Section~\ref{sec:discussions}.

\subsection{Selection of Stars} 
\label{sec:ranL}
\noindent
A series of runs were undertaken to appraise the influence of the mode
of selection of stars from the N-body realization of the cluster. The
kinematic information of these chosen stars then form the input file
for the inverse algorithm. One way to pick the stars from the N-body
data set is to choose a certain number of stars at a
random. Alternatively, we could sort the stars by their luminosity and
accept the $N$ brightest stars to form an input data file of size $N$
(here $N$ is any integer). We seek to explore the effect of these
varying methods of building up the input data for CHASSIS.

In Figure~\ref{fig:ranL_hyades}, the enclosed mass and velocity
dispersion profiles of the Hyades cluster, as predicted by the code
are compared to the known (N-body) cluster configuration. The functions
are obtained from runs done with input data files of diverse sizes,
formed by choosing the constituent stars either randomly or by
luminosity, from model H. The results show that when the stars are
chosen by their luminosity, the match between the estimated and N-body
quantities is in general better than when the choice is random. In
fact Figure~\ref{fig:ranL_hyades} indicates that CHASSIS does a good
job in reproducing the cluster profiles when the stars are chosen by
their luminosity. The results from the run done with the larger number
of stars (199 stars) is particularly good. Even the run done with the
much smaller number of stars (17) offers results that are consistent
with the N-body mass profiles, within the error bars, which are indeed
rather larger. However, the overlap between the velocity dispersion
profile obtained from the latter run is not so good. One reason for
this general lack of agreement between the estimated velocity
dispersion and the N-body dispersion profile is the shortcoming in the
very method of extraction of the dispersion profile from the N-body
data set. This is explained in details in Section~\ref{sec:conclu}. 

\section{Results from model A}
\label{sec:arches}
\noindent
N-body simulations of the Arches cluster gave us our model A. We have
performed a number of runs with data selected from this model and fed
into CHASSIS, in order to obtain the characteristics of the Arches
cluster. Runs were implemented to evaluate the effect of changing the
size of the input data set and of changing the manner of selecting
stars from model A. The N-body simulation of this cluster has been
performed at much lower radii compared to that of the Hyades cluster.
One interesting difference between the two clusters is that the Arches
cluster is identified as highly aspheric in the N-body
calculations. Isotropy in velocity space is also a worse assumption
for model A than model H. We have confined ourselves to the very
central core of the Arches cluster (radius $\leq$ 0.15pc) where the
effects of asymmetry in phase space are not too prominent
(Figure~\ref{fig:anisotropy}). 
 
Figure~\ref{fig:mass_arch} shows the mass within the central 0.093pc
(the virial radius approximately), estimated by the code for input
data sets of various sizes. This predicted value is compared to the
N-body value of 24,522M$_\odot$ of the enclosed mass. 

The effect of the process of selecting the stars from the N-body
realization of the Arches cluster is displayed in
Figure~\ref{fig:ranL_arches}. The figure shows the cluster profiles
obtained from a number of runs which are performed with differently
sized input files that were constructed by picking stars randomly from
model A. From these runs, it is clear that the inverse algorithm
over-estimated the enclosed mass at larger radii. When the stars are
chosen in decreasing order of their luminosity, the runs show better
match with the N-body mass profiles, within error bars. The
compatibility with the known velocity dispersion profiles was likewise
poorer for these runs.

\section{Discussions}
\label{sec:discussions}
\noindent
The aim of this paper was to advance the case of using CHASSIS to
characterize stellar clusters. We proceeded in this exercise by first
introducing the reader to the salient features of the inverse
algorithm CHASSIS. This was followed by testing the code with
simulated kinematical data for two different clusters, Hyades and
Arches. 

The signatures of the clusters, as known via the N-body simulations
were compared to the characteristics recovered by the code. The degree
of overlap between the simulated cluster configurations and the
predicted ones is expected to shed light on the applicability of the
algorithm. The relevant cluster attributes were judged to be the
enclosed mass and the line-of-sight projected velocity dispersion
profiles.

We tested the robustness of the algorithm to variations in the initial
guess for the cluster characteristics (see
Figure~\ref{fig:alfexp_massvel}). Results were compared from runs done
with the steepness of the initial density profile varying from
$\alpha=-1.5$ to -2.5 and -3.5 at a fixed choice of the initial
DF. The recovered mass profiles agreed with each other very well
within error bars (as well as with the N-body profile). Similarly
promising results were obtained from runs done with a fixed initial
density profile and a DF which varied in steepness ($\beta$ was
changed from 5.5 to 6.5 and 7.5). In regard to the velocity dispersion
profiles, it is seen that the overall match is good though when seen
over short length scales, the degree of overlap is not as good as it
is in the case of the mass profiles. This is in fact a feature that is
consistently borne by all the presented velocity dispersion profiles.

This discrepancy owes more to the way the known dispersion is drawn
out from the N-body simulations of the clusters than on the
calculation of the dispersion by CHASSIS. The projected radii from the
N-body data sets are binned and the velocities of the stars in each
such bin is calculated. (For model H, stars with velocities lower
than 1kms$^{-1}$ were used in this calculation while for the Arches
cluster, stars in each bin with velocities lower than 20kms$^{-1}$
were used). This discrete distribution of velocity dispersions is
plotted in Figure~\ref{fig:dispersions} as crosses.  This distribution
is ``optimally'' smoothed to represent the velocity dispersion of the
cluster. By ``optimal'' smoothing is meant smoothing done (using the
SuperMongo software) with a filter size which if exceeded renders the
plot less jagged but does not alter the overall shape. This smoothed
run of the dispersion with $r_p$ is what the calculated (by CHASSIS)
dispersion profile is compared to. This method of extracting the
dispersion profile from the N-body results is indeed {\it ad hoc} to
some extent but the methodology is adopted consistently; consequently,
for the purposes of comparison, it is good enough.

There is another technical reasons why in general the match between the
calculated and known velocity dispersion profiles is less good than
that between the enclosed mass profiles. The mass profile of the
relevant cluster is estimated from its density via
Equation~\ref{eqn:encmass} while velocity dispersion is calculated
using Equation~\ref{eqn:DF_disp}. Since the latter equation involves
two summations unlike only one in Equation~\ref{eqn:encmass}, it is
expected that any error in the estimation of the DF will imply greater
deviation in the evaluated dispersion value than any error in the
estimated density will contribute to the mass profile. 

The quantities that the inverse algorithm directly offers are the
density profile and the equilibrium distribution function of the set
of stars for which kinematical data is used as input to the code.
These quantities were then used to calculate the mass and dispersion
profiles, under the assumptions of sphericity and isotropy in velocity
space. These assumptions were essentially made to ease our
calculations but are not necessarily realistic constraints on
clusters. For example, we recognize the Arches cluster as highly
aspheric from the N-body simulations. In order to be able to predict
stable results with CHASSIS, we confined our analysis to radii within
which the degree of asphericity and anisotropy is markedly low.

The way to ensure that the set of stars which is used as input to
CHASSIS is closer to a spherical configuration is by using the
kinematical data of the more massive stars. These stars form a subset
in the cluster that is relatively more spherical. It is after all the
tidal field of the Galaxy, which when included in the N-body
simulations, causes the cluster to elongate in one direction.  This
field has a weaker effect on the heavier than the less massive stars
in the cluster since the more massive stars tend to reside closer to
the center of the cluster owing to mass segregation. This is confirmed
by Figure~\ref{fig:compare}: the left panel in this figure represents
the run of the ratio of the transverse to the radial velocity
component of 200 stars chosen from model A. In one case the stars are
chosen by luminosity (results depicted in unfilled circles) while in
the other case they are chosen randomly (filled triangles). The
distribution of the ratio $v_{\mu}/v_z$ with $r_{p}$ is more centrally
concentrated when the more massive (i.e. roughly speaking the brighter
stars) are included in the sample -- there are comparatively more
triangles at higher radii. The right panel of the same figure shows
that the effect of the tidal field of the Galaxy is more pronounced in
the sample constructed via random selection of the stars. The value of
apparent position is indexed by a number in the data set representing
the sample; the ratio $r_p/z$ is plotted against this number for the
two samples. The distributions of the ratios are smoothed by the same
amount in both cases and show that the sample of stars constructed by
random choice is less spherical than when the choice is
by mass.

In this context, let us reiterate that the choice on the basis of
stellar mass roughly resembles the choice based on luminosity. In
fact, a few runs were carried out with the $N$ most massive stars and
the profiles obtained from these runs corroborated the results already
mentioned.

Thus, if the input set of $N$ stars ($N$ is a natural number) is
formed by choosing the $N$ most massive stars within a certain radius,
instead of choosing $N$ stars randomly, the assumption of sphericity
is more closely obeyed and the enclosed mass profile as recovered by
CHASSIS is expected to be more compatible with the N-body profile. In
fact this is true even interior to the radial cutoffs that we have
imposed in our selection of the stars from the N-body outputs (of 10pc
and 0.15pc for the Hyades and Arches clusters respectively), as
verified by Figure~\ref{fig:compare}.

If the set of stars used as input to the algorithm form an aspherical
configuration, then it is to be expected that our code will
over-estimate the mass, (see Figure~\ref{fig:overestimate}). This is
what is seen from runs done with input data sets that are formed by
choosing stars randomly. This is evident in the lower panels in
Figure~\ref{fig:ranL_hyades} and Figure~\ref{fig:ranL_arches};

We have also explored the effect of changing the number of data points
($N$) which are used as input to the code. Naively, one may conclude
that a larger number of data points invariably improves the result in
the sense that the error-bars diminish in size. The larger the data
set, the narrower is the region around the maximum of the likelihood
function in the multi-dimensional energy-radius space. Consequently,
the wandering of the Metropolis algorithm in its ``equilibrium'' phase
is smaller, thus bringing down the extent of the errors. Albeit it is
expected that increasing the number of input data points is soon going
to lead to nearly stable results. 

It is noticed that until the number of stars in the input data set is
about 80, the simulated and predicted quantities are closer for the
Hyades cluster than when the Arches cluster is used (see
Figure~\ref{fig:ranL_hyades} and Figure~\ref{fig:ranL_arches}). One
reason for this is that the Arches cluster is known from the N-body
simulations to be much more aspheric than the Hyades. The degree of
anisotropy in velocity space is also greater in Arches. As stated in
Section~\ref{sec:N-body}, even in the central 0.15pc, the ratios of the
different components of the velocity (and spatial) dispersion, when
sought as functions of radius, undulate about a mean of unity. The
same quantities for the Hyades cluster are less non-linear inside the
chosen radial cutoff for model H (about 10pc). Thus, for the Arches
cluster, stability in the results is attained at a higher value of $N$
(at about $N=80$) than when input data is chosen from model H for
which this number is 17.  This is evident in the comparison of
Figures~\ref{fig:mass_hyades} and ~\ref{fig:mass_arch}. 

Our code is rather powerful in the sense that it can spot the answers
even when $N$ is quite low. We propose that this potency is a direct
outcome of implementing the Metropolis algorithm as the
optimizer. Metropolis is a highly competent algorithm and is more
effective than when the maxima in the likelihood function is sought by
brute force as in other non-parametric projects that have used an
isotropic distribution function \citep{davidsaha,merrittremblay}. When
the likelihood is highly non-linear, such a method is not strong
enough to recover the maximal region of the likelihood unless the
number of input data points is sufficiently high to adequately
constrain the likelihood. Even when Metropolis is used, we find that
for a model such as model H, we need to have around 100 data points to
constrain the enclosed mass at the half-mass radius within a range of
about 120M$_\odot$, i.e. fractional errors of about 20$\%$.

The approximation of isotropy, as used in the algorithm, aids in
simplifying the calculations but needs to be improved via the
inclusion of a two-integral distribution function in the code. This is
planned as a future exercise.

\section{Conclusions}
\label{sec:conclu}
\noindent
We have carried out analysis of two N-body clusters using the inverse
algorithm, CHASSIS. Our code aims to simultaneously recover the
potential and the distribution function of the stars whose kinematic
details are fed in as input. The global maxima in the likelihood
function is sought by the Metropolis algorithm. The cluster profiles
directly offered by the algorithm are used to calculate the
line-of-sight projected velocity dispersion and enclosed mass profiles
of the cluster.

We looked into the robustness of the code with respect to change in
the used initial parameters. The nearly concurrent results reflect
that the code does indeed withstand variations in the input
configurations very well (see Figure~\ref{fig:alfexp_massvel}). 

When input data was used from the model that mimics the Hyades
cluster, the enclosed mass and the velocity dispersion profiles
predicted by the code were consistent with the simulated profiles for
the Hyades cluster as long as the number of kinematic data points used
as input in the code is at least 17 though the error bars that
accompany the results from this run are very large. For the Arches
cluster, an input data set with at least as many as 63 stars is
required to yield a mass profile that is consistent with the N-body
profile within the error bars though the velocity dispersion profile
from this run falls short of the predicted N-body dispersions
(Figure~\ref{fig:ranL_arches}).  The input data set needs to contain
kinematic information of at least 80 stars to result in both mass and
dispersion profiles that are compatible with the known profile, within
the error bars.

The overlap between the quantities suggested by the code and the N-body
profiles is better when the input model follows the assumption of
phase space isotropy closely, i.e. within about 10pc for the Hyades
and 0.15pc for the Arches cluster.

When the input file comprises the brightest stars, the predictions of
the code match the cluster profiles better than when the stars are
picked randomly. Roughly speaking, the brightest stars tend to
follow the trend of the most massive stars and are segregated toward
the cluster center while the dimmer stars are more prominent in the
cluster halo. Thus, the observer who is likely to look toward the
center stands a greater chance of picking out those stars for which
the code predicts the correct results. Consequently, CHASSIS can claim
to be a propitious tool for the analysis of data of stellar clusters.

Increasing the number of stars in the input file does not affect the
results beyond decreasing the size of the error bars; for a given
model, there is a minimum value of this number, which when exceeded
implies stable results. This value is low when the cluster is very
nearly spherical and isotropic. For the Arches cluster this numerical
value is around 80 while it is as low as about 17 for the Hyades
cluster though the run done with these 17 stars yields very large
error bars; at the half-mass radius of the N-body model for the Hyades
cluster, the mass enclosed varies in the range of about 500M$_\odot$
to 900M$_\odot$. (As $N$ increases, the error bars reduce in size
considerably, so that when $N\sim$200, the mass within the half-mass
radius ranges from about 582M$_\odot$ to about 586M$_\odot$). We
believe that is it our sophisticated optimization routine that
utilizes the Metropolis algorithm that helps us to identify the
correct answers even with quite small input data sets. The search for
the range of $N$ that corresponds to consistent results can be fully
automatized.

Our numerical experiments offer confidence in the applicability of
CHASSIS as a tool to analyze stellar clusters. The algorithm has been
shown to be successful when the cluster is spherical and isotropic in
velocity space.  The assumption of isotropy is an ingredient in the
code which is under review.

CHASSIS has been used previously to estimate the mass of the central
black hole in the Galaxy (Paper~I) and in this paper, to
characterize open stellar clusters. In its present form, it is put
forward as a useful tool to understand kinematic data of other stellar
systems such as clusters.

\section*{Acknowledgments}
\noindent
DC would like to acknowledge Dr. P. Saha who is the joint author of
CHASSIS. SPZ was supported by the Royal Dutch Academy of Sciences
(KNAW) and by the Netherlands Organization of Scientific Research
(NWO). We would also like to thank the anonymous referee whose careful
scrutiny helped to improve the paper considerably.

\clearpage
\begin{figure}
\plotone{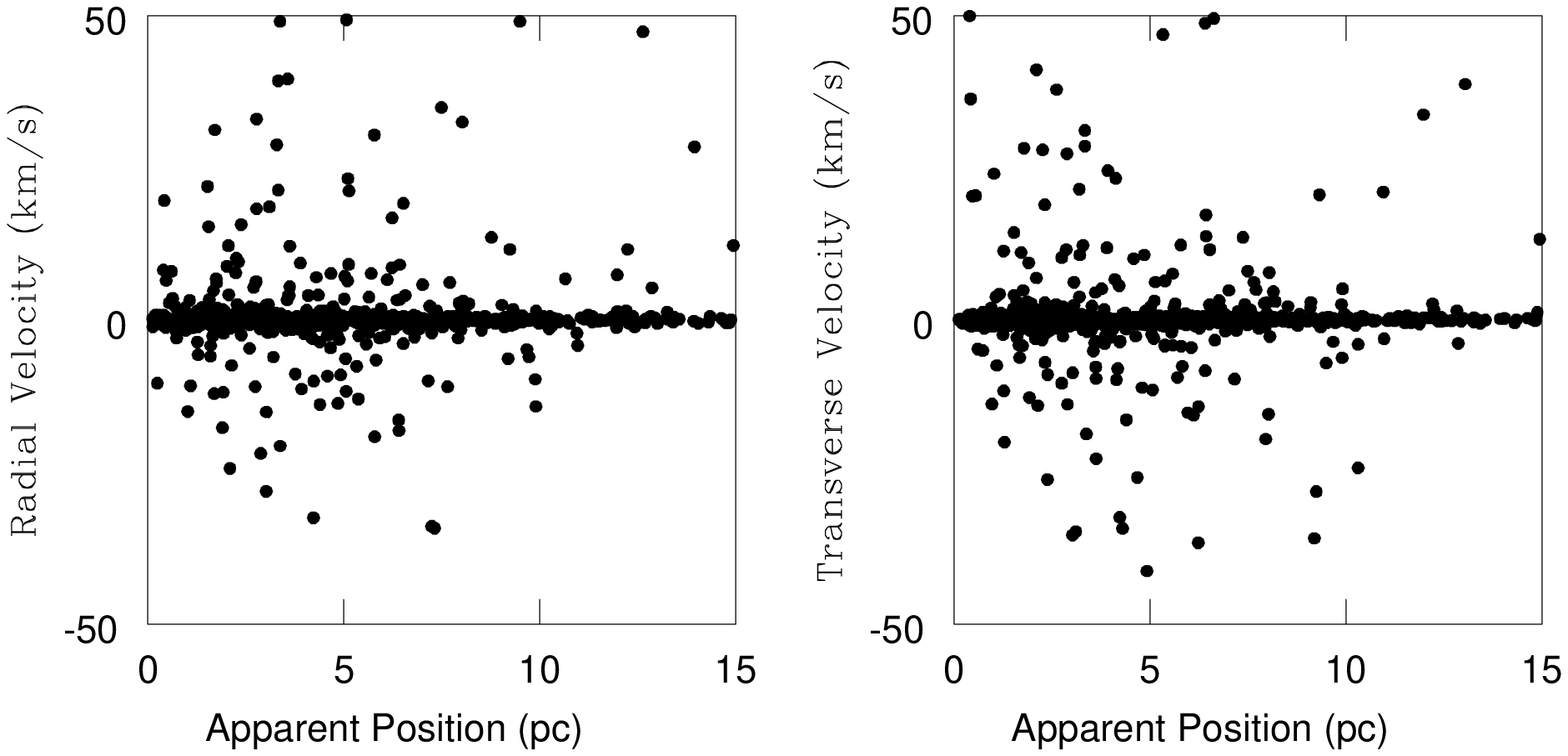}
\end{figure}
\begin{figure}
\plotone{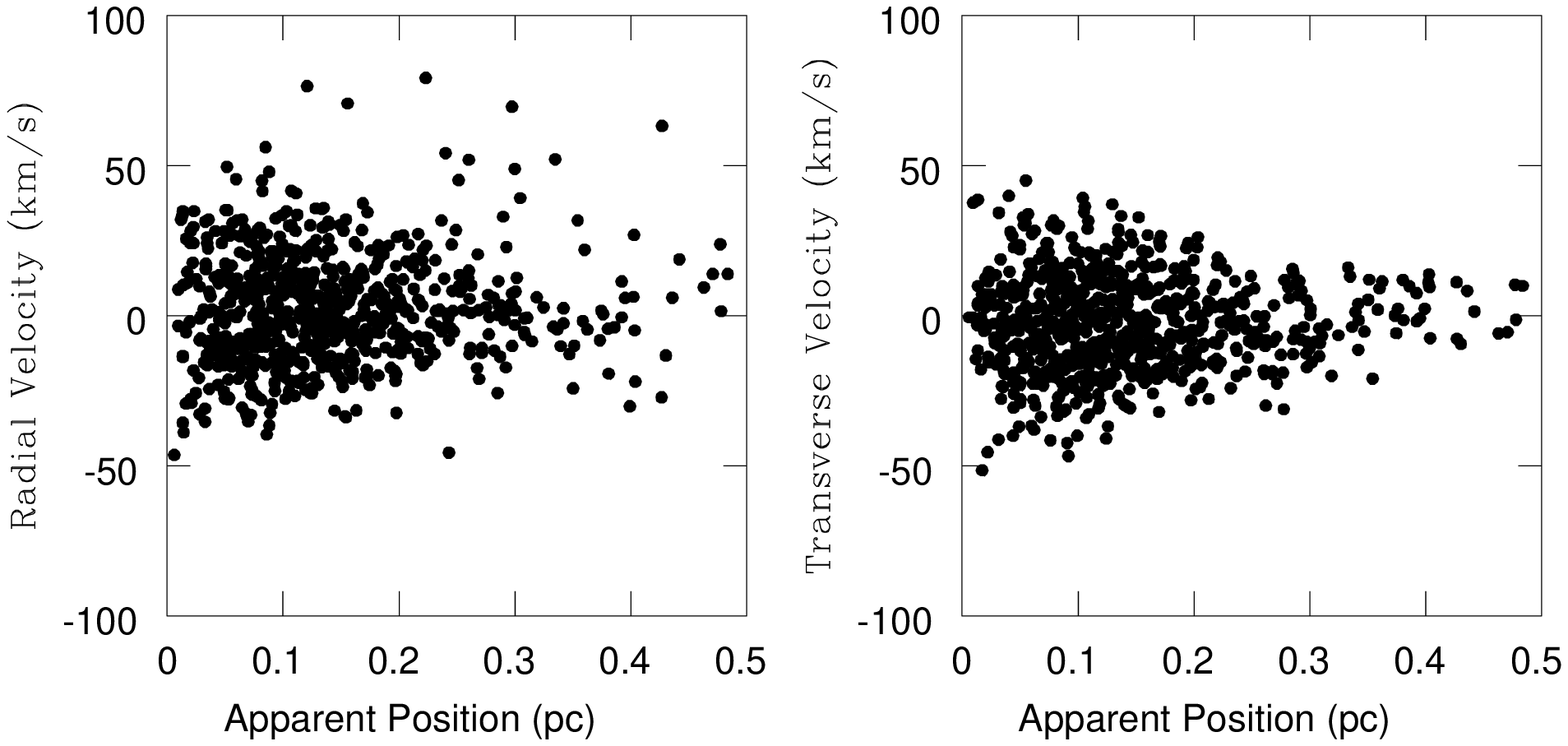}
\end{figure}
\begin{figure}
\plotone{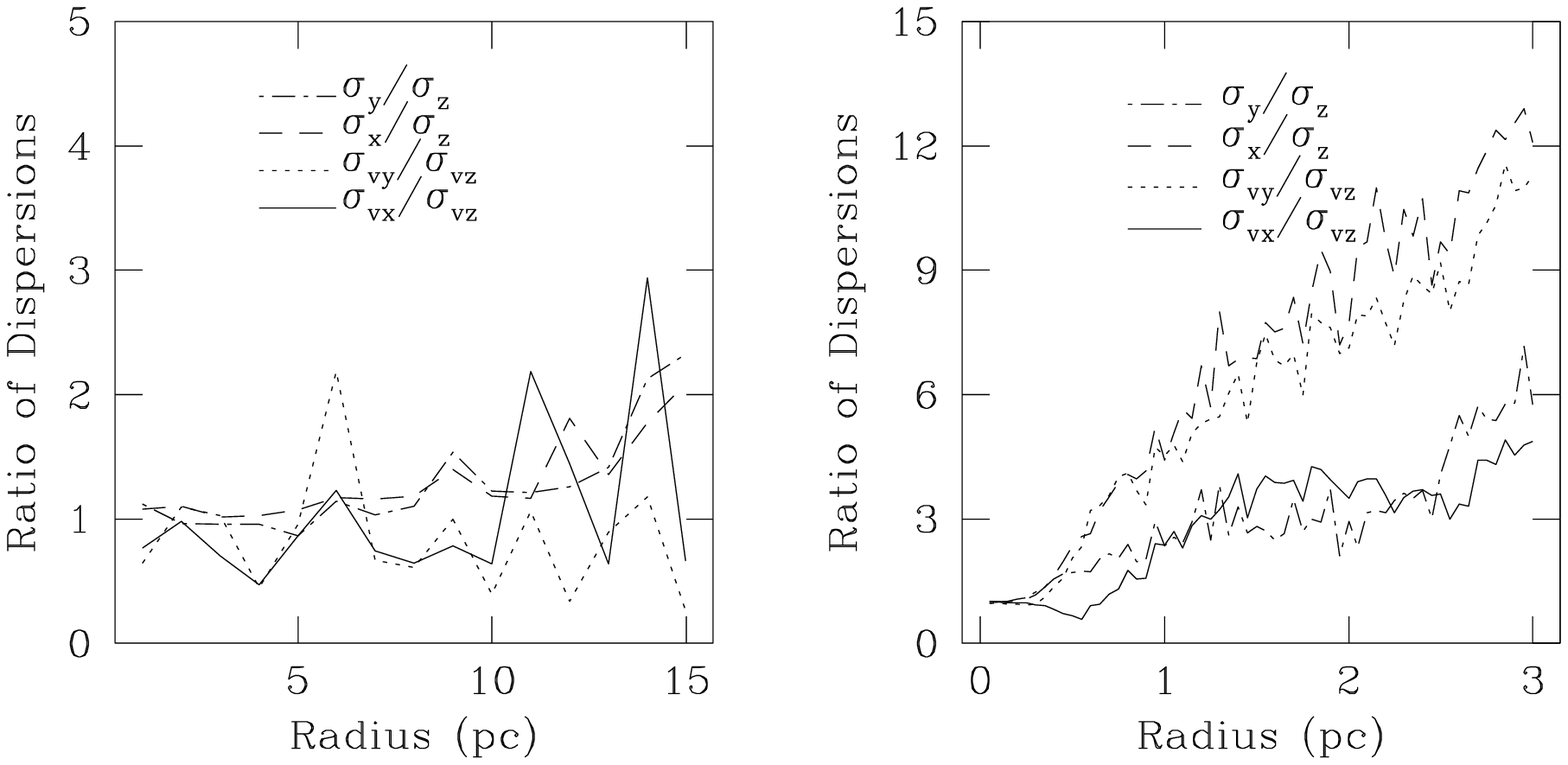}
\end{figure}
\begin{figure}
\plotone{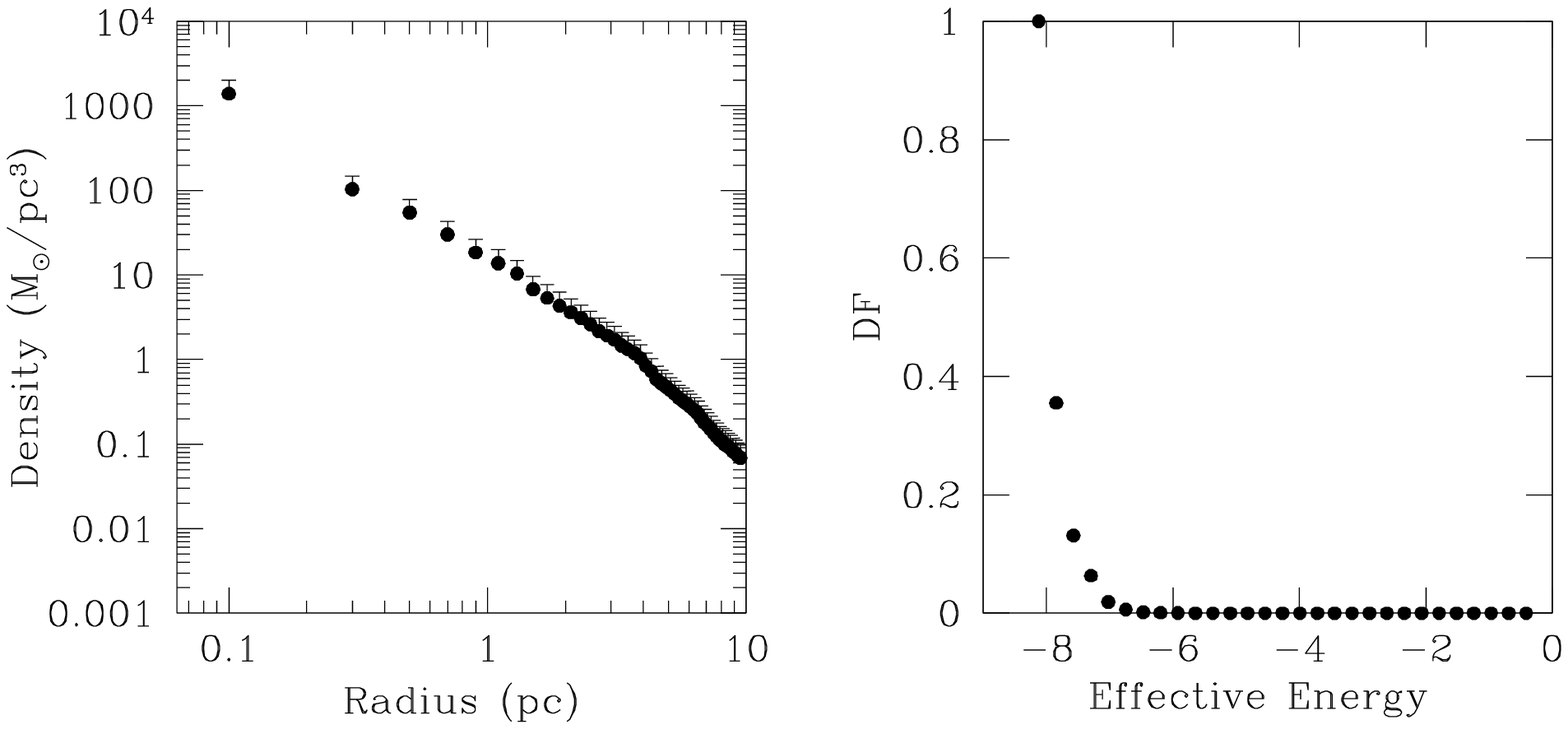}
\end{figure}
\begin{figure}
\plotone{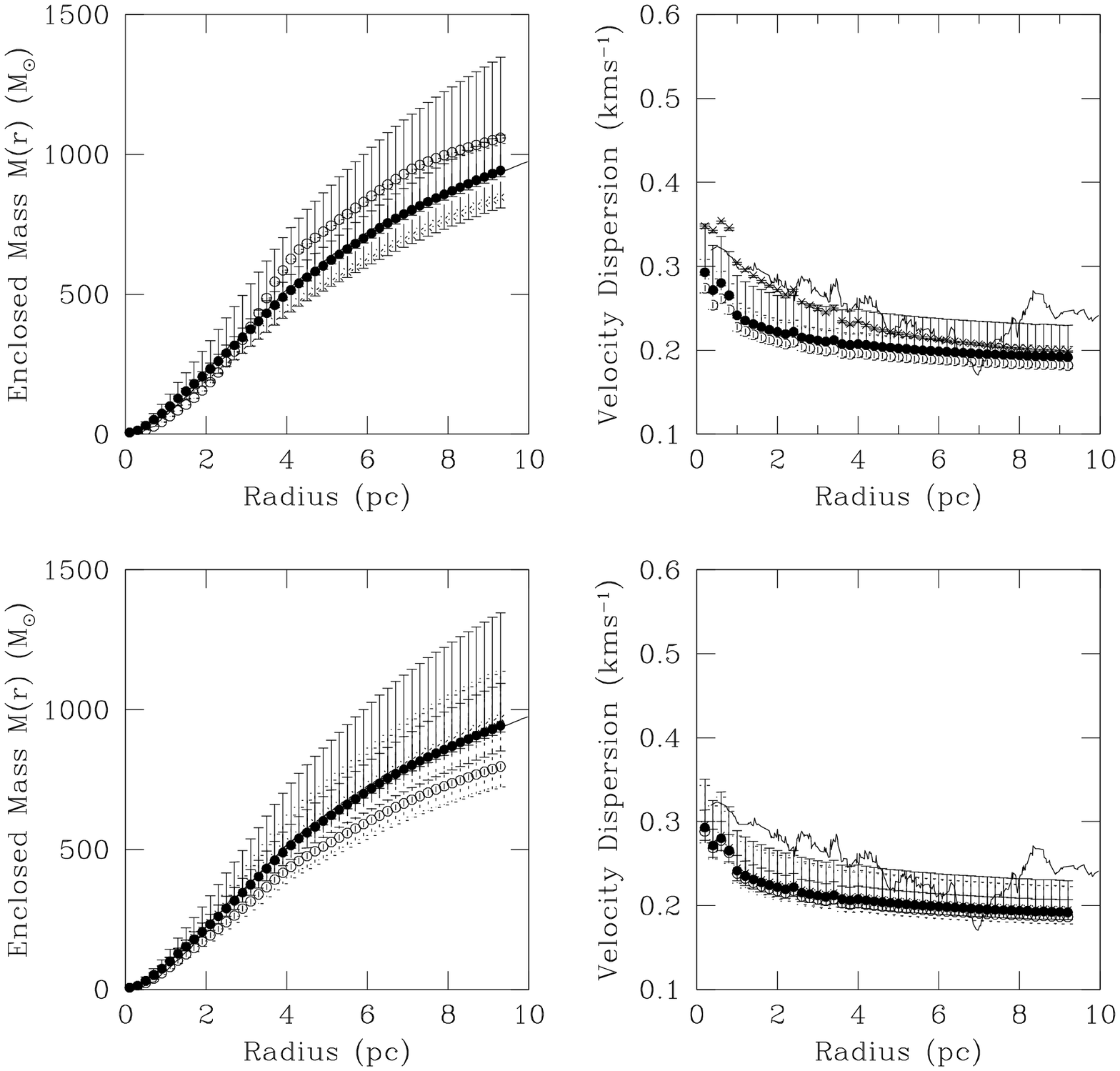}
\end{figure}
\begin{figure}
\plotone{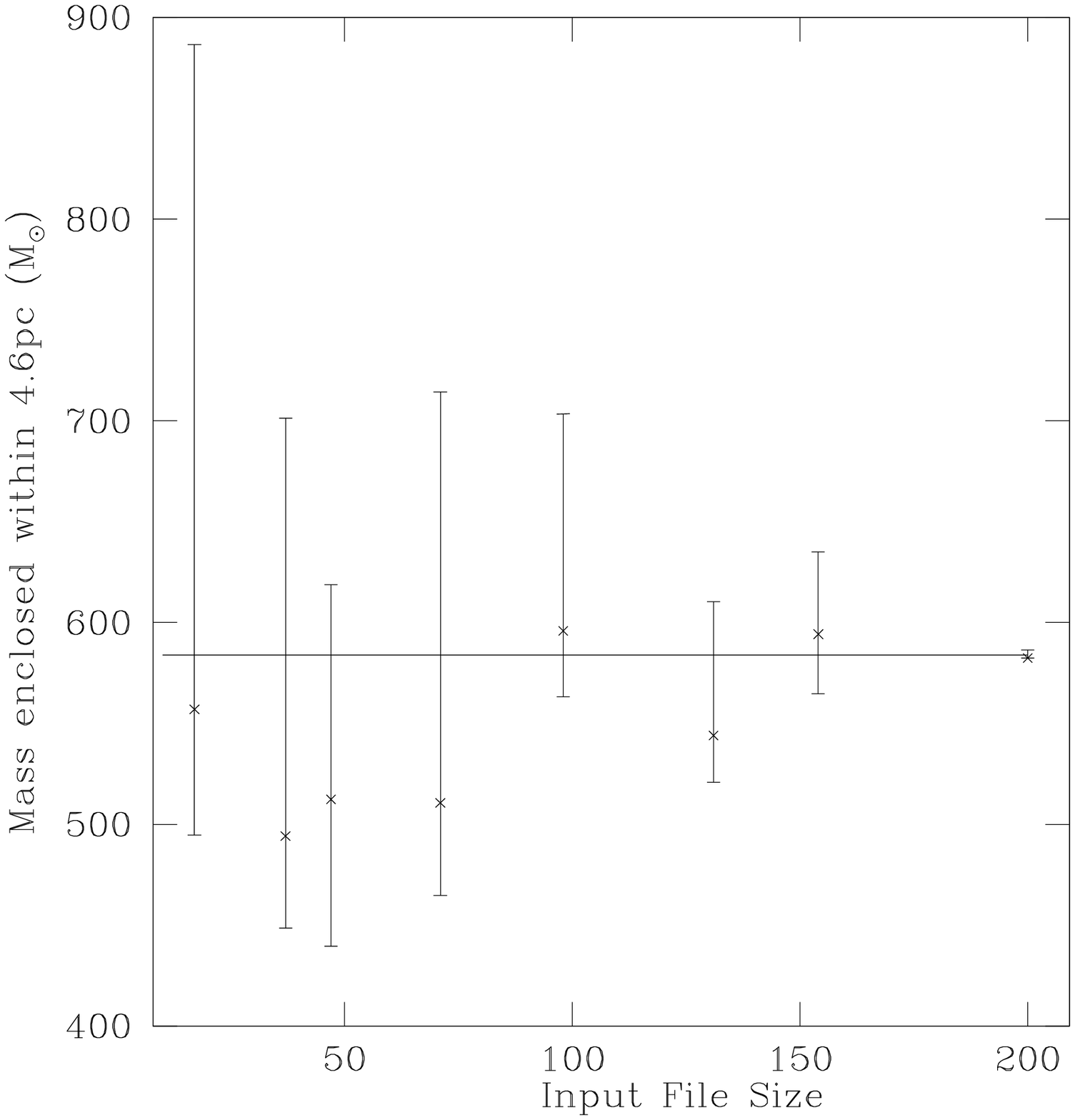}
\end{figure}
\begin{figure}
\plotone{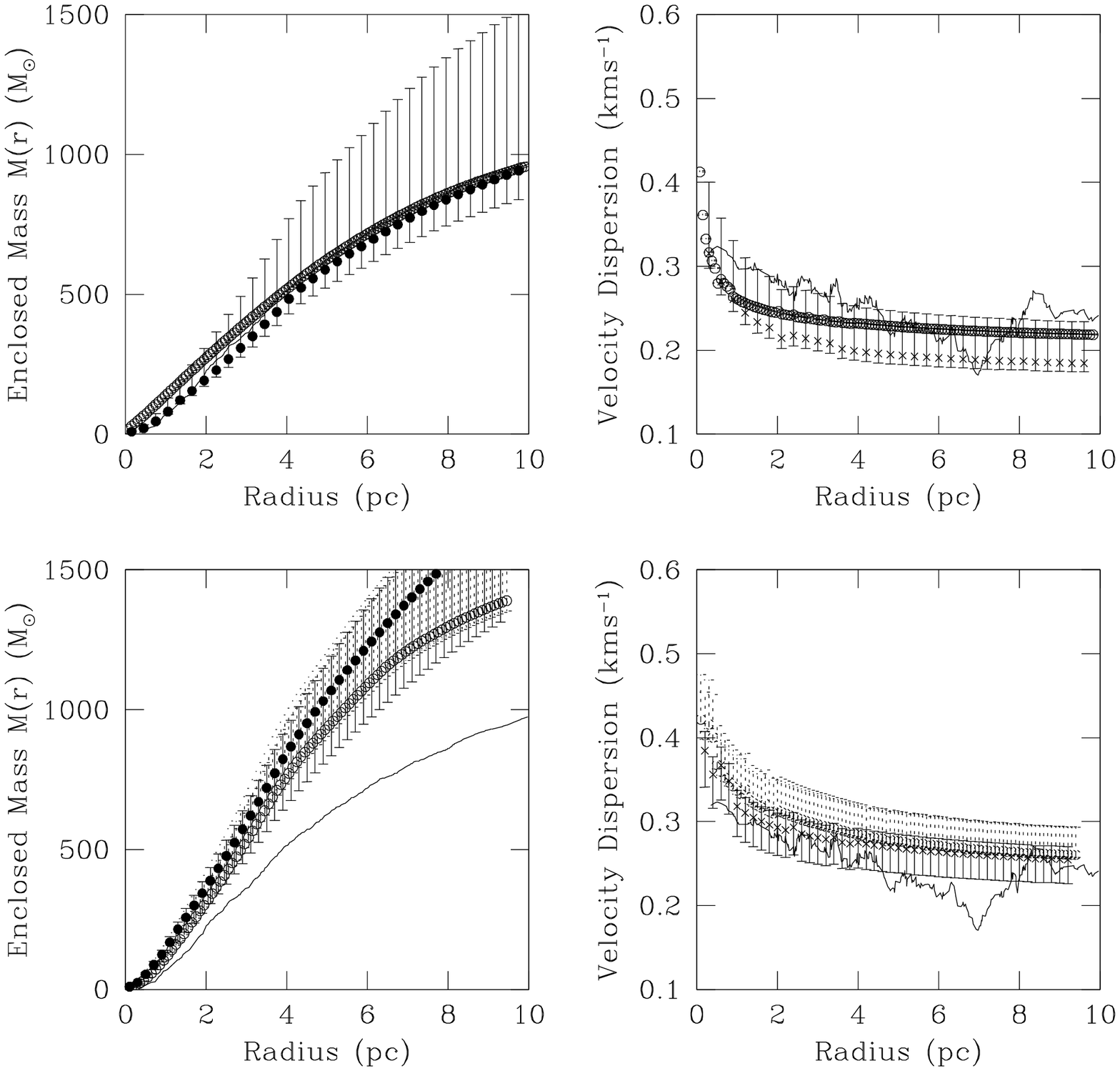}
\end{figure}
\begin{figure}
\plotone{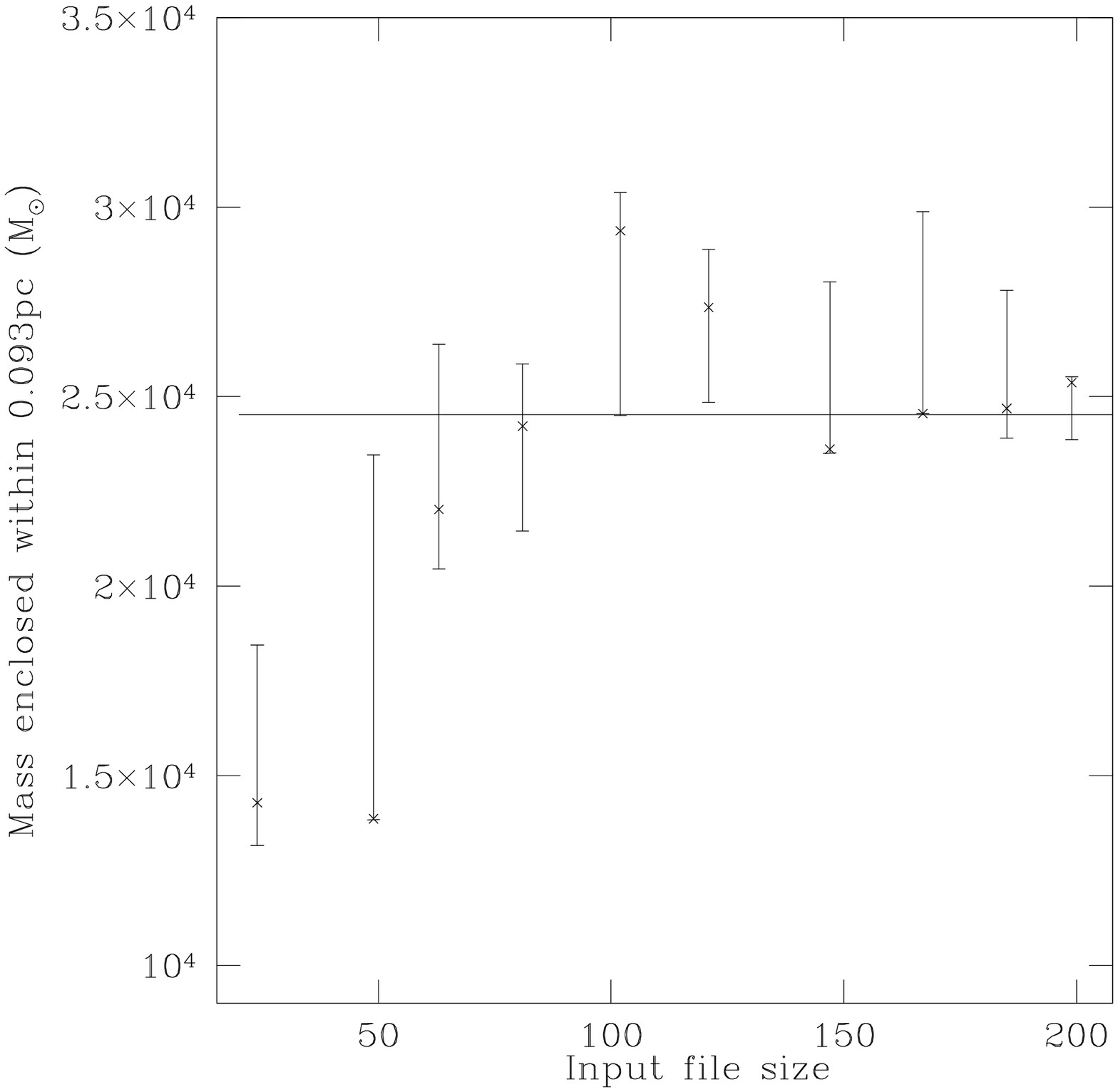}
\end{figure}

\begin{figure}
\plotone{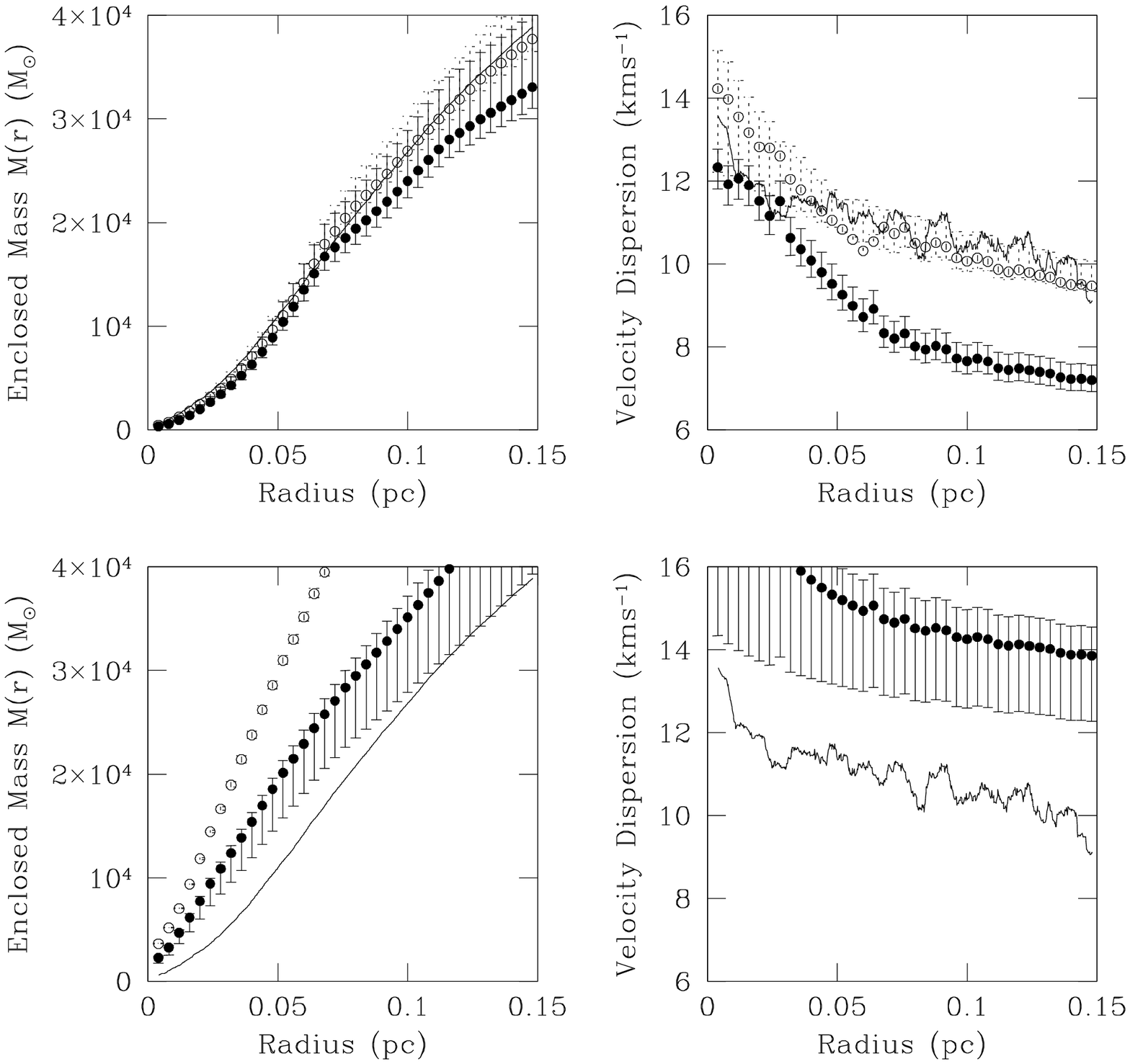}
\end{figure}
\begin{figure}
\plotone{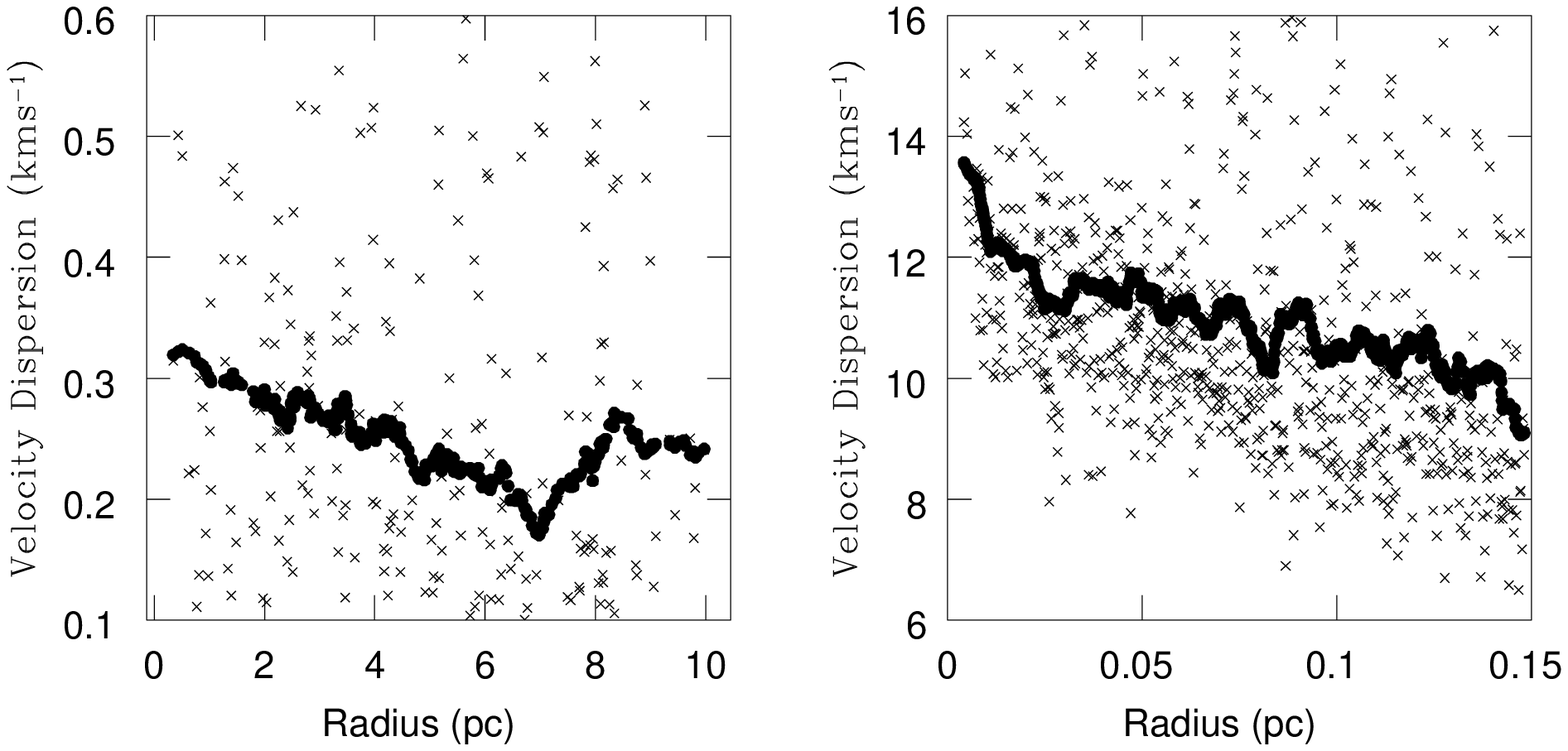}
\end{figure}
\begin{figure}
\plotone{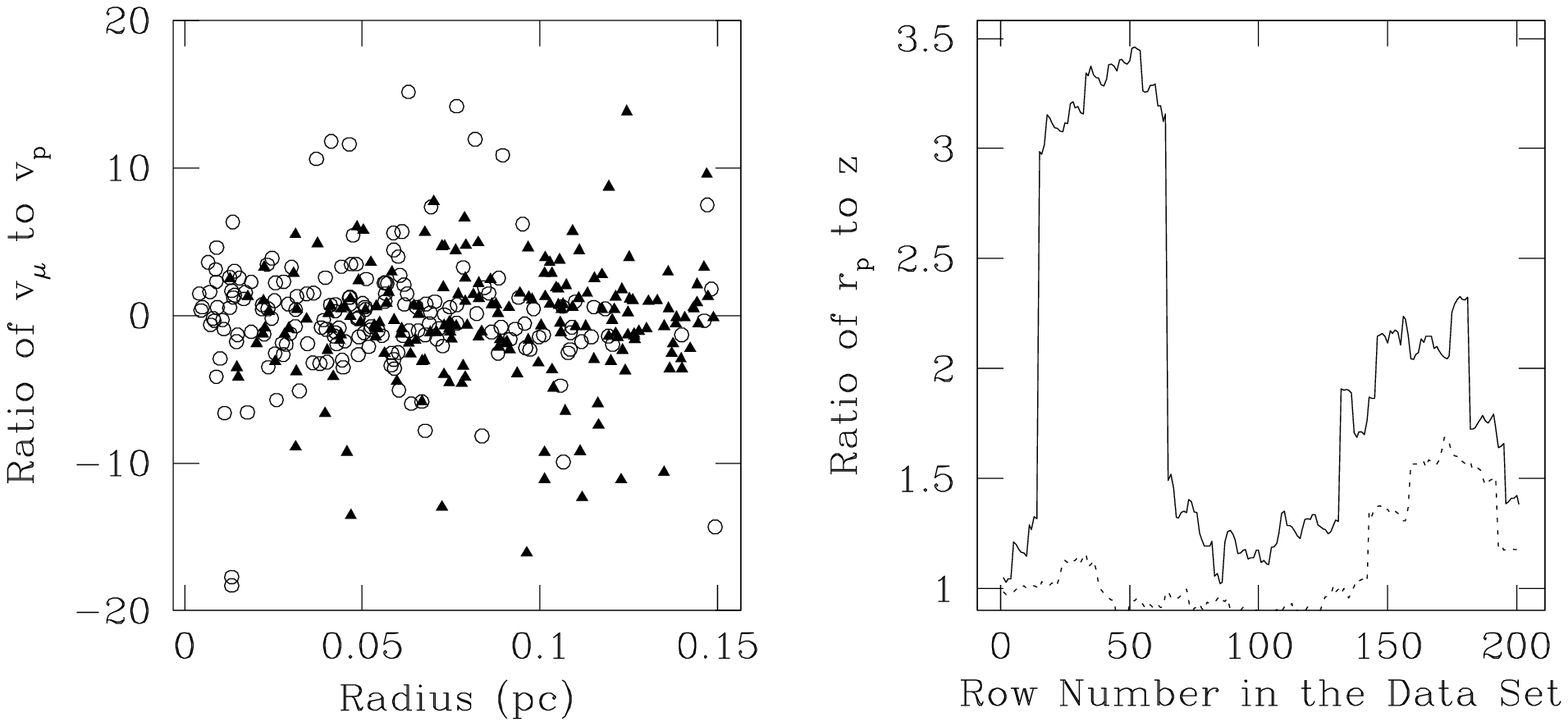}
\end{figure}
\begin{figure}
\plotone{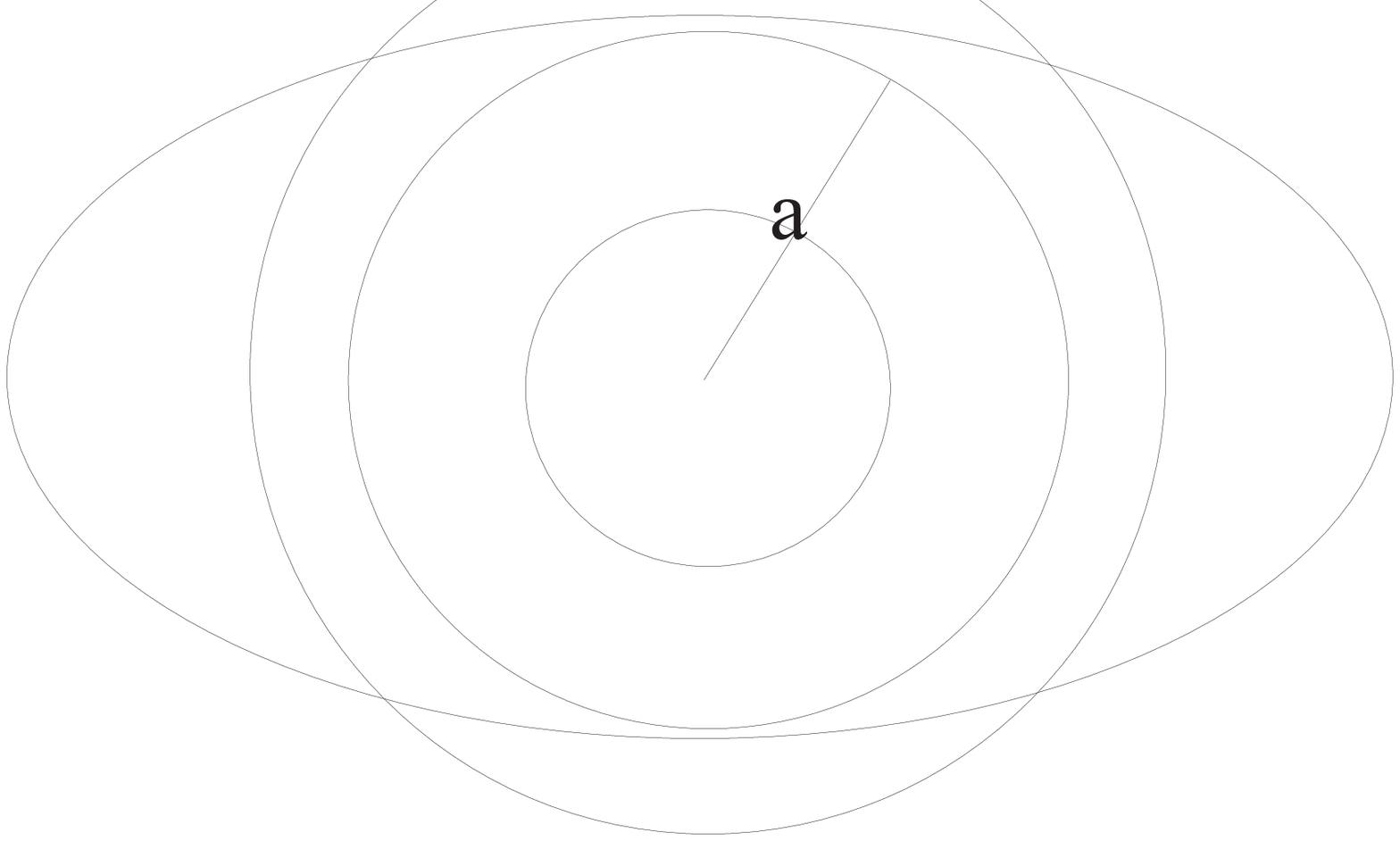}
\end{figure}

\clearpage

\figcaption[dc.fig1.ps]
{Plots of radial (left panel) and transverse (right panel) 
velocities of stars in model H against apparent positions. The data is
obtained from N-body calculations described in the
Section~\ref{sec:N-body}. To construct this plot, we have chosen every
3$^{rd}$ star in the N-body representation of this cluster.
\label{fig:data200}}

\figcaption[dc.fig2.ps]
{Plots of radial and transverse velocities of stars in model A,
against apparent position. This N-body cluster has 55,224 stars in it
but only stars at pre-fixed intervals in projected radius have been
included in the plot.
\label{fig:data64}}

\figcaption[dc.fig3.ps]
{Ratio of the $x$ and $y$ components of the velocity (and spatial)
dispersion, $\sigma_{vx}$ and $\sigma_{vy}$ (and $\sigma_{x}$ and
$\sigma_{y}$) to the $z$ component of the velocity (an spatial)
dispersion, $\sigma_{vz}$ (and $\sigma_z$).
\label{fig:anisotropy}}

\figcaption[dc.fig4.ps]
{Figure showing the density (left panel) and equilibrium stellar
 distribution function (right panel) of the Hyades cluster, as
recovered from a run done with the initial configuration given by
$\alpha=-1.5$ and $\beta=5.5$, using an input file of 37 randomly
picked stars from model H. The error bars are at the $16\%$ and the
$84\%$ levels.
\label{fig:dfden_H}}

\figcaption[dc.fig5.ps]
{Profiles of the Hyades cluster, as estimated by CHASSIS,
from runs performed with various initial configurations. The left
panels display the enclosed mass profiles while the velocity
dispersion profiles recovered from the different runs are shown in the
right panels. In the upper panels, the index $\beta$ is held a
constant (at 5.5) and $\alpha$ is changed. The profile for
$\alpha=-1.5$ is marked by error bars in solid lines with superimposed
filled circles at the 50$\%$ level, those for $\alpha=-2.5$ are in
dotted lines with open circles and for $\alpha=-3.5$ errors are in dashed
lines with crosses. In the lower panels the initial density profile is
maintained the same ($\alpha=-1.5$) but the initial choice of the DF
varies in steepness. $\beta=5.5, 6.5 \& 7.5$ are marked by error bars
in solid, dotted and dashed lines, superimposed with filled circles,
open circles and crosses respectively. The results obtained from the
calculations of the data describing model H are shown in solid black
lines. The error bars are at the $16\%$ and the $84\%$ levels.  In the
dispersion profiles, the error bar at any radius is calculated as the
difference between the dispersion value corresponding to the value of
density at the $84\%$ level and at the $16\%$ level. The input for the
algorithm is the set of the 37 brightest stars in model H.
\label{fig:alfexp_massvel}}

\figcaption[dc.fig6.ps] 
{Figure showing the estimates of mass enclosed
within the inner 4.6pc in model H (the half-mass radius), indicated by
CHASSIS, from runs done with input data of a varying number of stars,
selected by their luminosity. The mass estimates are shown with
superimposed error bars (at the 16$\%$ and 84$\%$ levels). The N-body
calculations imply a mass of about 584M$_\odot$ of the cluster. This value
is represented by a solid line.
\label{fig:mass_hyades}}

\figcaption[dc.fig7.ps]
{Figure showing the enclosed mass and velocity dispersion profiles 
of Hyades, as claimed by CHASSIS from runs done with input data that
has either been picked randomly (lower panels) or according to the
luminosity of stars (upper panels). When the stars are sorted by their
luminosity, the $N$ most luminous among them are selected from model
H, where $N$ is an integer. For $N$=17, the error bars are in solid
lines superimposed with filled circles (to mark the 50$\%$ level)
while they are shown in dotted lines with open circles for $N=199$.
When the stars are picked randomly, the match is comparatively
worse. Two values of $N$ are used for these runs; profiles are shown
with errors in solid lines superimposed with filled circles for $N=37$
and dotted lines with open circles for $N=71$. The curve in solid line
represents the profiles indicated by the N-body calculations.
\label{fig:ranL_hyades}}

\figcaption[dc.fig8.ps] 
{Similar to Figure~\ref{fig:mass_hyades}
except that here the mass enclosed within the inner 0.093pc of the
Arches cluster (half-mass radius) is displayed. The mass estimates are
from runs done with input files of various sizes, constructed by
picking stars by their luminosity from model A. The N-body estimate of
the mass within this radius is about 24,522M$_\odot$. This value is
represented by the solid black line.
\label{fig:mass_arch}}

\figcaption[dc.fig9.ps]
{Similar to Figure~\ref{fig:ranL_hyades} except that here the 
profiles of the Arches cluster are shown. When the input file size $N$
is 63 the error bars are in solid lines and the symbol type at the
50$\%$ mark is filled circle while these are dotted lines and open
circle respectively for $N=185$. In the lower panels, for $N=17$ the
error bars are in solid lines, superimposed with filled circles while
the same are in dotted lines and open circles for $N=200$. 
\label{fig:ranL_arches}}

\figcaption[dc.fig10.ps]
{Figure showing the discrete distribution of velocity dispersion
values with radius. At each (projected) radial bin, the raw stellar
velocities from the output of the N-body simulations are used to
calculate the velocity dispersion in that bin. These are plotted as
crosses against projected radius. An optimally smoothed curve is fit
to this distribution of points. By ``optimal'' is implied that size of
the smoothing filter which if exceeded helps to reduce the unevenness
of the plot but retains the overall shape of it. This velocity
dispersion plot obtained from the N-body model is compared to the
LOS projected dispersion estimated by CHASSIS.
\label{fig:dispersions}}

\figcaption[dc.fig11.ps] 
{Figure showing the effects of choosing stars randomly as
distinguished from building a sample by selecting the most massive
(i.e. roughly speaking, the brightest) stars. In the left panel, the
ratio of the transverse ($v_{\mu}$) to the radial ($v_p$) velocity of
200 stars in the sample, is plotted against projected radius. When the
choice is random, the result is shown in filled triangles while
selection by luminosity is represented by open circles. 
The position of a value of $r_p$ in the data set is indexed by a row
number; the ratio $r_p/z$ is plotted against this number in the right
panel. The results are equally smoothed for the two samples. The
distribution of the ratio is in broken lines when the selection is by
mass and is in solid lines when the choice is random.
\label{fig:compare}}

\figcaption[dc.fig12.ps] {Figure to show why we should expect our
algorithm to overestimate the enclosed mass in an aspheric cluster, at
higher radii. The ellipse represents the 2-D projection of a perfectly
oblate cluster. Within radii $R\leq{a}$ the enclosed mass is correctly
calculated by the code. A fraction of any annulus at $R\:<\:a$ lies
outside the cluster. Now, the code uses the density in the shell at
any radius, to spread the mass uniformly throughout that
annulus. Thus, the mass estimated to lie inside the annulus at radii
$>\:a$ is in excess of the actual mass enclosed within this radius.
\label{fig:overestimate}}

\end{document}